\documentclass[journal,twoside,web]{ieeecolor}
\usepackage{generic}
\usepackage{cite}
\usepackage{amsmath,amssymb,amsfonts}
\usepackage{graphicx}
\usepackage{textcomp}
\usepackage{xcolor}
\usepackage{multirow}

\usepackage{lipsum}

\usepackage{hyperref}
\hypersetup{
    colorlinks=true,
    linkcolor=black,
    urlcolor=blue
    }
\def\BibTeX{{\rm B\kern-.05em{\sc i\kern-.025em b}\kern-.08em
    T\kern-.1667em\lower.7ex\hbox{E}\kern-.125emX}}
\begin{document}

\pagenumbering{gobble}
\newpage 
\clearpage

\thispagestyle{empty}
\onecolumn
\begin{table}[ht]
\centering
\begin{tabular}{p{\linewidth}}
© 2020 IEEE. Personal use of this material is permitted.  Permission from IEEE must be obtained for all other uses, in any current or future media, including reprinting/republishing this material for advertising or promotional purposes, creating new collective works, for resale or redistribution to servers or lists, or reuse of any copyrighted component of this work in other works.
Accepted for publication in IEEE Transactions on Electron Devices (Special Section on ESSDERC 2020). DOI: 10.1109/TED.2020.3019019

\end{tabular}

\end{table}

\newpage
\twocolumn
\pagenumbering{arabic}
\clearpage
\setcounter{page}{1}

\title{Generalized Constant Current Method for\\ Determining MOSFET Threshold Voltage}
\author{Matthias Bucher, \IEEEmembership{Member, IEEE}, Nikolaos Makris, and Loukas Chevas
\thanks{This work was partly supported under Project INNOVATION-EL-Crete (MIS 5002772).}
\thanks{M. Bucher, N. Makris and L. Chevas are with the School of Electrical and Computer Engineering, Technical University of Crete, 73100 Chania, Greece. (e-mail: bucher@electronics.tuc.gr).}
}

\maketitle

\begin{abstract}
A novel method for extracting threshold voltage and substrate effect parameters of MOSFETs with constant current bias at all levels of inversion is presented. This generalized constant-current (GCC) method  exploits the charge-based model of MOSFETs to extract threshold voltage and other substrate-effect related parameters. The method is applicable over a wide range of current throughout weak and moderate inversion and to some extent in strong inversion. This method is particularly useful when applied for MOSFETs presenting edge conduction effect (subthreshold hump) in CMOS processes using Shallow Trench Isolation (STI).
\end{abstract}


\section{Introduction}
\label{sec:introduction}
Constant current (CC) methods for the extraction of MOSFET threshold voltage are very convenient due to their simplicity, when compared to classical methods using typically extrapolation of drain current in strong inversion. However, CC methods often use a somewhat arbitrary current criterion to determine threshold voltage, e.g., $I_{Th}\,=\,100nA\,\cdot ({W}/{L})$ \cite{b1},\cite{b2},\cite{b3}. This current criterion is then applied to measured transfer characteristics ($I_D-V_G$) of MOSFETs, from which threshold voltage may be determined as a function of back-bias voltage. CC methods are simple and robust, but the resulting parameters depend on the current criterion used. In \cite{b4},\cite{b5}, the current criterion is established such that the transistor operates in the middle of moderate inversion. The aim of this work is to present a generalization of the current criterion to any level of current, from weak through moderate and to strong inversion. As will be shown, using this generalized current criterion, threshold voltage can be determined from transfer characteristics at any level of current. Hence, the method will be termed generalized constant current (GCC) method. The procedure relies on the charge-based model formulation \cite{b6},\cite{b7}, and enables extracting threshold voltage and other substrate effect related parameters. Note that the adjusted-constant current (ACC) method \cite{b5} allows for the determination of threshold voltage from linear to fully saturated operating conditions. Linear mode operation brings benefits in minimizing the impact of second-order effects on the extracted parameters, such as velocity saturation (VS), channel length modulation (CLM), and drain induced barrier lowering (DIBL). The GCC method presented here concentrates on saturated operating conditions as a generalization of \cite{b4}. As will be discussed, the GCC method is preferably applied at weak and moderate levels of inversion, and hence the impact of 
vertical field, series resistance and VS effects are kept minimal.

A particular field of application of the GCC method is the characterization of edge conduction (or "subthreshold hump") effect in MOSFETs. The latter effect often arises in technologies using shallow-trench isolation (STI) schemes. As will be shown, the GCC method is a powerful tool to determine parameters for the edge conduction effect.  

\section{Generalized Constant Current Method in Weak and Moderate Inversion}
\label{sec:GCCM}
The generalized constant current method to extract threshold voltage is intricately linked with the charge-based model approach \cite{b6},\cite{b7}. In the latter, drain current is written,
\begin{equation}
I_{D} = I_{spec}\left(i_f - i_r\right) = I_{spec}\left(q_{s}^2 + q_{s} - q_{d}^2 - q_{d} \right) \label{i-q}
\end{equation}
where $i_{f(r)}=q_{s(d)}^2 + q_{s(d)}$ are the normalized forward(reverse) current components depending on the inversion charge densities $q_{s(d)}$ at source(drain), respectively. The same relationship can also be reversed as $q_{s(d)}=\sqrt{1/4+i_{f(r)}}-1/2$. The specific current $I_{spec}$ is defined as, 
\begin{equation}
I_{spec}\,=\,I_0\,\cdot\left({W}/{L}\right)\,=\,2\,n\,U_T^{2}\,\mu\,C_{ox}\,\left({W}/{L}\right) \label{ispec}
\end{equation}
which may be obtained from measurement as in \cite{b4},\cite{b5}. In the above, $n$ is the slope factor, mobility is $\mu$, and $C_{ox}=\epsilon_{ox}/T_{ox}$ the gate capacitance depending on oxide thickness $T_{ox}$. The voltage-charge relationship is expressed as \cite{b6},
\begin{figure*}[tbph]
	\centerline{%
		\begin{tabular}{ccc}
			\includegraphics[scale=0.175]{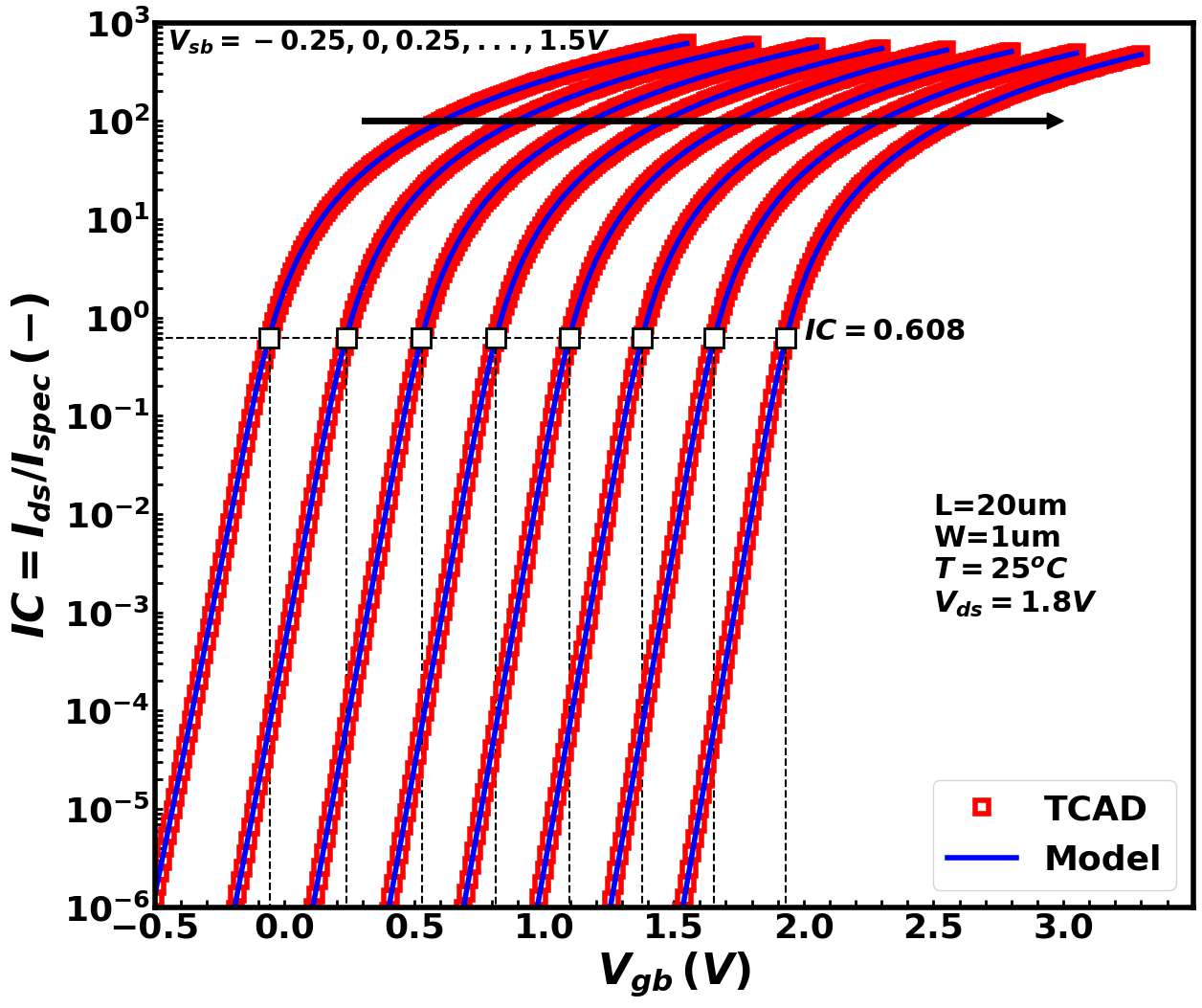}  & \includegraphics[scale=0.175]{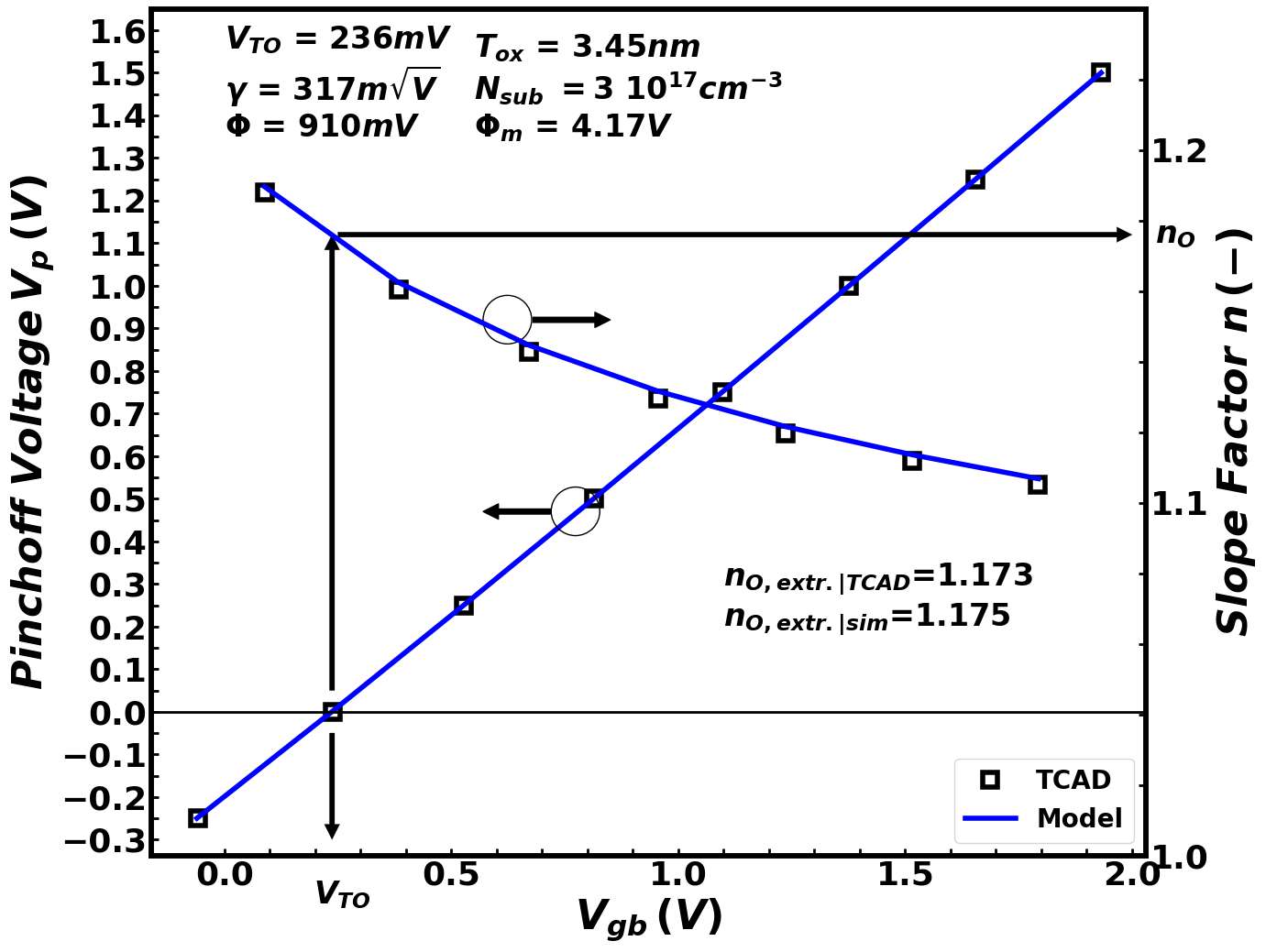}  & 
			\includegraphics[scale=0.175]{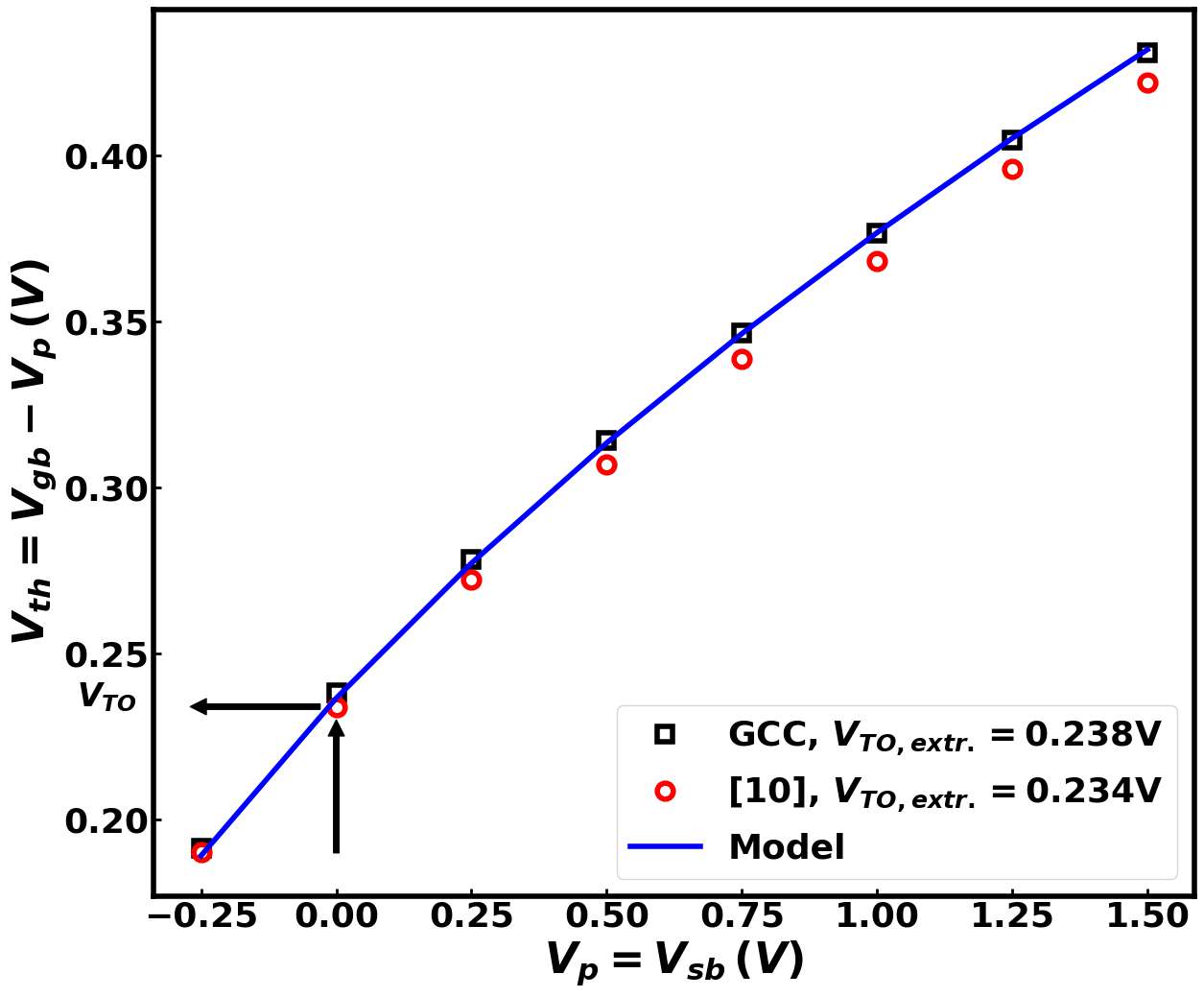}
	\end{tabular}} \caption{Principle of the generalized constant current (GCC) method using $IC=0.608$	in moderate inversion. Transfer characteristics $I_D$ vs. $V_G$ with applied constant current (left), extracted pinchoff voltage $V_P$ and slope factor $n$ vs. $V_G$ (center) of a wide/long n-MOSFET in saturation. $V_{TO}$ is determined as $V_G|_{V_P=V_S=0}$. The GCC method is applied similarly to TCAD data (markers) and charge-based model (lines) either using constant, non field-dependent mobility. The corresponding $V_{TH}$ vs. $V_S$ (right) shows close agreement with $\partial (G_m/I_D)/ \partial V_G$ method \cite{b10}.}
\end{figure*}
\begin{equation}
(V_P-V_{S(D)})/{U_T}=v_p-v_{s(d)}=2q_{s(d)}+ln(q_{s(d)}) \label{q-v}
\end{equation}
where voltages $v_p={V_P}/{U_T}$ and $v_{s(d)}={V_{S(D)}}/{U_T}$ may be normalized to the thermal voltage $U_T = kT/q$. The pinch-off voltage $V_P$ is a function of the gate voltage $V_G$ \cite{b6}-\cite{b8},  
\begin{equation}
\begin{split}
V_P&=V_G-V_{FB}-\Psi_0-\gamma\left[\sqrt{ V_G-V_{FB} + \left( \frac{\gamma}{2} \right)^2} -\frac{\gamma}{2}\right]\\
&\approxeq (V_G-V_{TO})/{n(V_G)}
\label{vpvg}
\end{split}
\end{equation}
where $V_{FB}$ is the flat-band voltage, $\gamma=\sqrt{2q\epsilon_{si} N_{sub}}/C_{ox}$ is the substrate effect factor, $\Psi_0\simeq2\Phi_F + 2...3U_T$ is slightly above twice the Fermi potential $\Phi_F=U_Tln{(N_{sub}/n_i)}$ \cite{b6}, $N_{sub}$ and $n_i$ are substrate doping and intrinsic concentrations, and $V_{TO}=V_{FB}+\Psi_0+\gamma\sqrt{\Psi_0}$ is the threshold voltage. 

The inverse function $V_G(V_P)$ \cite{b8},\cite{b9} is equally useful, 
\begin{equation}
V_G = V_{TO} + V_P + \gamma \left[ \sqrt{\Psi_0+V_P} - \sqrt{\Psi_0} \right] \label{vgvp}
\end{equation}
as it allows us to define both the gate threshold voltage $V_{TB} \equiv V_G|_{V_P=V_{S}}$ (referred to local substrate), as well as the commonly used threshold voltage $V_{TH}$ (referred to source),
\begin{equation}
V_{TH} \equiv V_G - V_P|_{V_P=V_{S}} = V_{TO} + \gamma \left[ \sqrt{\Psi_0+V_P} - \sqrt{\Psi_0} \right]
\label{vth}
\end{equation}

The slope factor $n$ is defined as \cite{b8},
\begin{equation}
n \equiv \left[\frac{\partial V_P}{\partial V_G}\right]^{-1} = 1 + 
\frac{\gamma}{2\sqrt{V_P+\Psi_0}} \label{n}
\end{equation}

The combination of \eqref{i-q}-\eqref{vpvg} and \eqref{n} defines the full charge-based expression of drain current in all regions of operation of the MOSFET. In saturation, when $i_r \ll i_f$, \eqref{i-q} reverts to $I_D \simeq I_{spec} \cdot i_f$, which can be also stated as,
\begin{equation}
IC=\frac{I_D|_{sat.}}{I_{spec}}=\frac{I_D|_{sat.}}{I_0\cdot \left(\frac{W}{L}\right)} \label{ic}
\end{equation}
where $IC$ is the inversion coefficient \cite{b7},  \cite{b8}, with $IC<0.1$, $0.1<IC<10$, and $IC>10$ defining weak, moderate, and strong inversion, respectively.

In the following, the GCC method to extract threshold voltage at any level of inversion will be established. A first interesting observation is that when current components due to drift $i_{drift} = q_s^2$ and diffusion $i_{diff} = q_s$ are equal (i.e. $q_s^2=q_s=1$), then from \eqref{i-q} we have $IC=2$, and from \eqref{q-v}, $v_p-v_s=2$, i.e. $v_p=v_s+2$. Hence, biasing the transistor at $IC=2$ allows us to obtain $V_P$ simply as an offset from $V_S$, namely, $V_P=V_S+2U_T$. We note at this point that for reasons of practicality, the slight bias dependence of specific current $I_{spec}$ (due to dependence of slope factor $n$ on $V_G$) is neglected: $I_0$ is assumed constant, and hence, a constant value of $IC$ also implies a constant current $I_D$.

\begin{table}[htbp]
\caption{Normalized Constant Current Levels, Pinchoff Voltage Offset, and Transconductance-to-Current Ratio}
\label{table}
\centering
\begin{tabular}{|c|c|c|c|}
\hline
\multicolumn{1}{|l|}{${IC}$} & \multicolumn{1}{l|}{${q_s}$} & \multicolumn{1}{l|}{${v_p-v_s}$} & \multicolumn{1}{l|}{${g_{ms}/i_d}$} \\
\multicolumn{1}{|c|}{${=q_s^2+q_s}$} & \multicolumn{1}{c|}{${=\sqrt{\frac{1}{4}+IC}-\frac{1}{2}}$} & \multicolumn{1}{c|}{${=2q_s+ln(q_s)}$} & \multicolumn{1}{c|}{${=\frac{1}{\frac{1}{2}+\sqrt{\frac{1}{4}+IC}} }$} \\ \hline
10 &     2.7 & 6.4 & 0.27      \\ 
2 &     1 & 2 & 0.5            \\ 
1 &     0.618 & 0.755 & 0.618    \\ 
0.608 &    0.426 & 0 & 0.701     \\ 
0.1 &     0.092 & -2.2 & 0.916    \\ 
 \hline
\end{tabular}
\end{table}

{When generalizing the above procedure to any value of $IC$, we can obtain the pinchoff voltage $V_P$ from \eqref{i-q} and \eqref{q-v} as a function of $IC$ and $V_S$,} 
\begin{equation}
V_P=V_S+U_T\left(2\left[\sqrt{\frac{1}{4}+IC}-\frac{1}{2}\right]+ln\left[\sqrt{\frac{1}{4}+IC}-\frac{1}{2}\right]\right) \label{vp-ic}
\end{equation}

Some practical values of \eqref{vp-ic} are listed in Table I. In particular, we note that the condition $V_P=V_S$, corresponding to $IC=0.608$, has been exploited in \cite{b4} and \cite{b5}. In \cite{b5}, extraction of threshold voltage has also been extended to the case of non-saturation.

The procedure is applied to TCAD data as well as to the charge-based model, using the same parameters (180 nm CMOS technology) quoted in Fig.~1. The charge-based model \eqref{i-q}-\eqref{vpvg} fits the TCAD data perfectly well. The procedure of determining threshold voltage and other substrate effect parameters is illustrated. A current criterion of $IC=0.608$
is applied to $I_D-V_G$ characteristics in saturation, with varying $V_S$. The intercepts with 
$IC=0.608$ occur at specific gate voltages, defining the pinchoff voltage characteristic as $V_P=V_S$ vs. $V_G$. From the latter, the slope factor $n$ is obtained by derivation 
via \eqref{n}. The threshold voltage is determined as $V_{TO} = V_G|_{V_P=0V}$, while the other substrate effect parameters $\Psi_0$ and $\gamma$ are obtained by a simultaneous fit of equations for $V_P$ \eqref{vpvg} and $n$ \eqref{n} to the measured (or TCAD simulated) characteristics. The corresponding $V_{TH}$ vs. $V_S$ characteristic shows a close agreement among the GCC and transconductance-to-current ratio change \cite{b10} methods.

\begin{figure*}[tbph]
	\centerline{%
		\begin{tabular}{ccc}
			\includegraphics[scale=0.175]{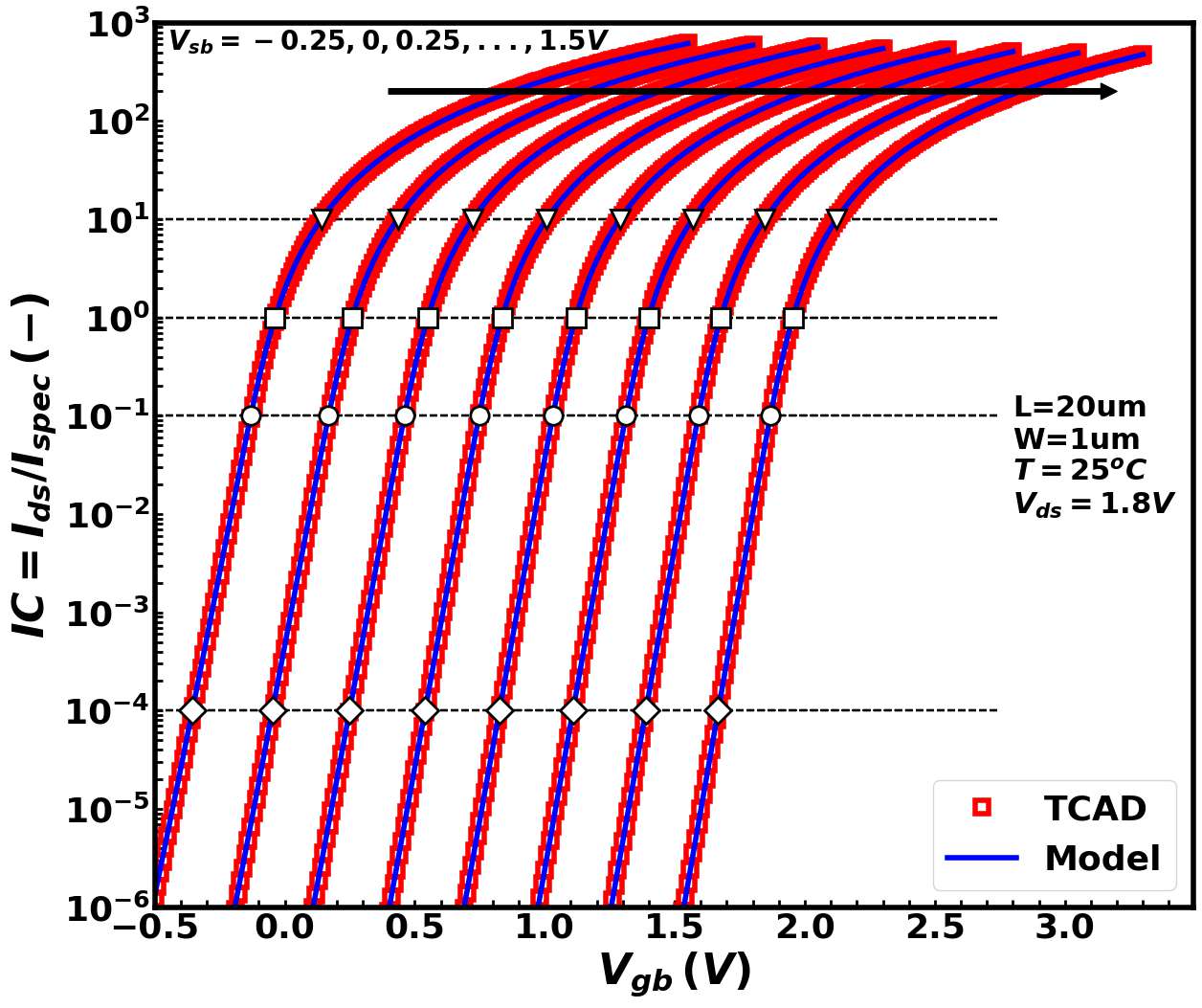}  & \includegraphics[scale=0.175]{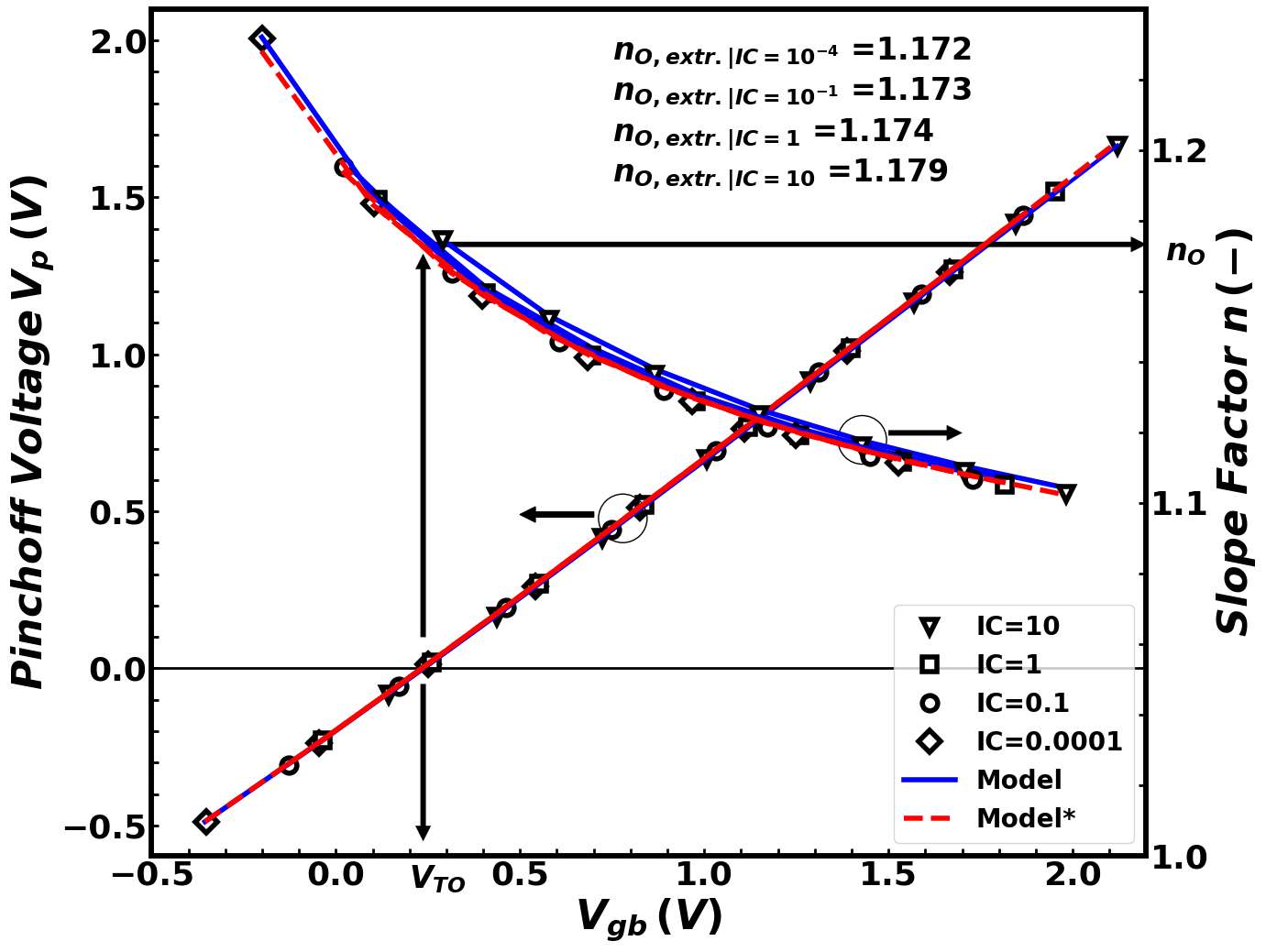}  & 
			\includegraphics[scale=0.175]{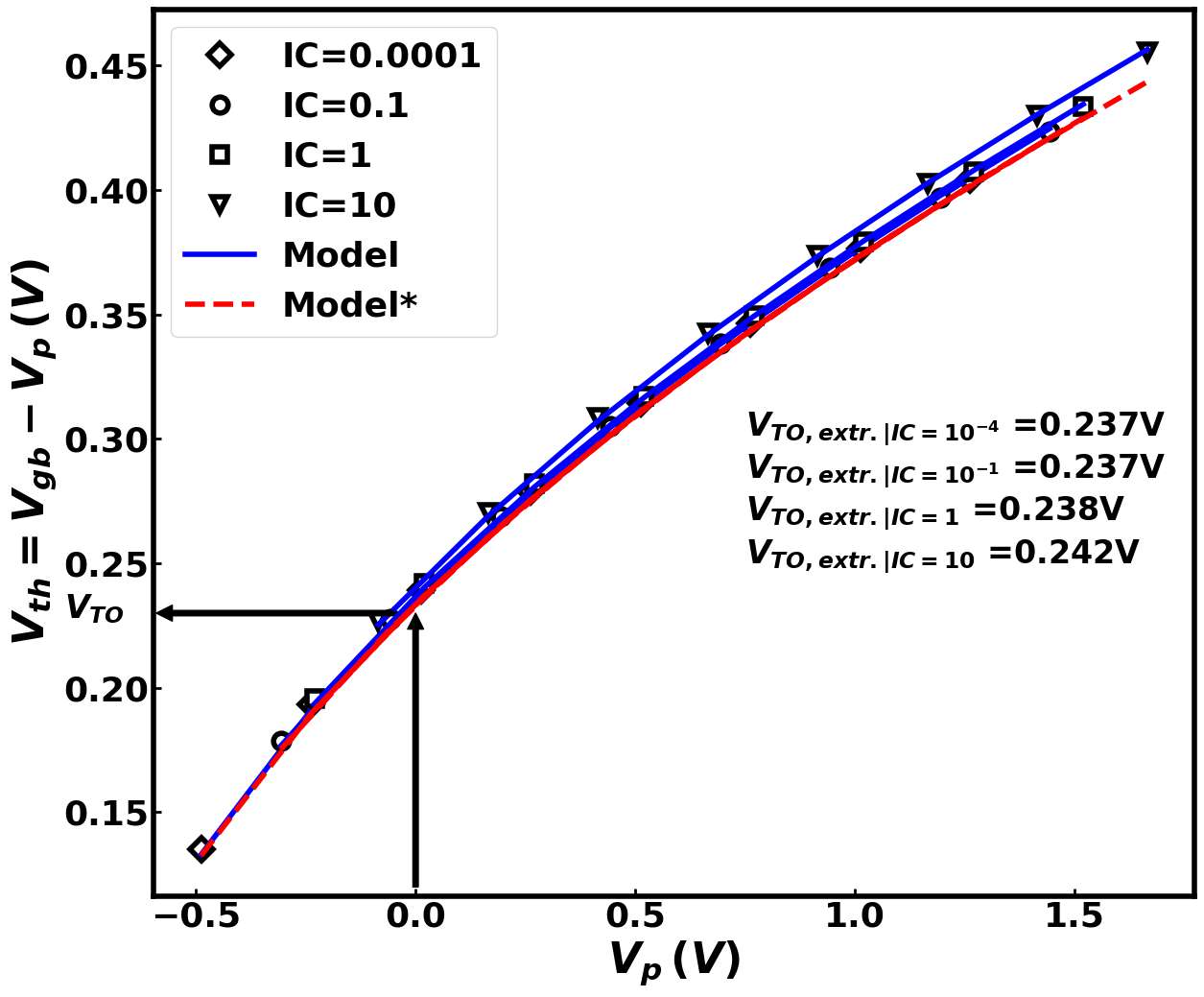}
	\end{tabular}} \caption{Generalized constant current (GCC) method, obtained at values of $IC = I_D/\left(I_0(W/L)\right)|_{I_0=cst.}$ ranging from very weak ($IC=10^{-4}$) to 
	onset of strong ($IC=10$) inversion. Transfer characteristics with applied constant current (left), extracted pinchoff voltage $V_P$ and slope factor $n$ vs. $V_G$ (center), and $V_{TH}$ vs. $V_P$ (right) of a wide/long n-MOSFET in saturation. $V_{TO}$ is determined as $V_G|_{V_P=0}$ consistently for a range of $IC$ spanning as much as five decades, resulting in a deviation of $V_{TO}$ of only 5 mV, while slope factor $n_0=n|_{V_P=0}$ varies by $<1\%$. The GCC method is applied in the same way to TCAD data (markers) and charge-based model (lines). A slight shift in $n$ vs. $V_G$ and $V_{TH}$ vs. $V_P$ is observed at higher levels of $IC$, consistently for TCAD data and model. This deviation is absent when the charge-based model (red dashed lines, Model*) also uses constant $I_0$.}
\end{figure*}

The generalized constant current method is illustrated in Fig.~2, showing the extraction procedure for a range of $IC$ values from very weak ($IC=10^{-4}$) to onset of strong ($IC=10$) inversion. The $V_P$ vs. $V_G$ characteristics obtained for the different levels of $IC$ are seen to practically overlap. The threshold voltage is again determined as $V_{TO} = V_G|_{V_P=0V}$, yielding a value of $V_{TO}$ within 5 mV over five decades of $IC$. The slope factor $n$, which is a derivative quantity, appears to be more sensitive, and presents a slight variation of $n_0=n|_{V_P=0}$ by less than 1 percent over the same range of $IC$. This may be attributed to the fact that the dependency of the slope factor $n$ on $V_G$ (affecting the drain current $I_D$ via $I_{spec}$) has been neglected in the extraction procedure. Such a practice is common among ACC \cite{b5}, transconductance-to-current ratio change \cite{b10}, transconductance-to-current ratio \cite{b11}, and GCC methods.

Indeed, when setting the slope factor to a constant value $n_0=n|_{V_G=V_{TO}}$ in $I_{spec}$ of the charge-based model, the GCC procedure yields the same slope factor, as indicated in Fig. 2 by dashed lines, coinciding for all levels of $IC$. Hence, the GCC method and the charge-based model are self-consistent. In other words, the GCC method is applicable in strong inversion if the dependence of the biasing current on the slope factor $n$ is anticipated. In practice, $IC$ values in moderate and weak inversion will be preferred, where the impact of vertical field, series resistance or VS effects are reduced \cite{b5},\cite{b10}. In general, the choice of the level of the current criterion may depend on device particularities, such as leakage, or testing conditions. 

To apply a (generalized) constant current criterion in the method as described above, one must first determine the specific current $I_{spec}$. This may be conveniently done from transconductance-to-current ratio at moderate levels of inversion. The normalized source and gate transconductance-to-current ratios $g_{ms}/{i_d}$ and $g_m/{i_d}$ in saturation are \cite{b5}\cite{b7},

\begin{figure}[tbph]
	\centerline{%
		\begin{tabular}{c}
			\includegraphics[scale=0.21]{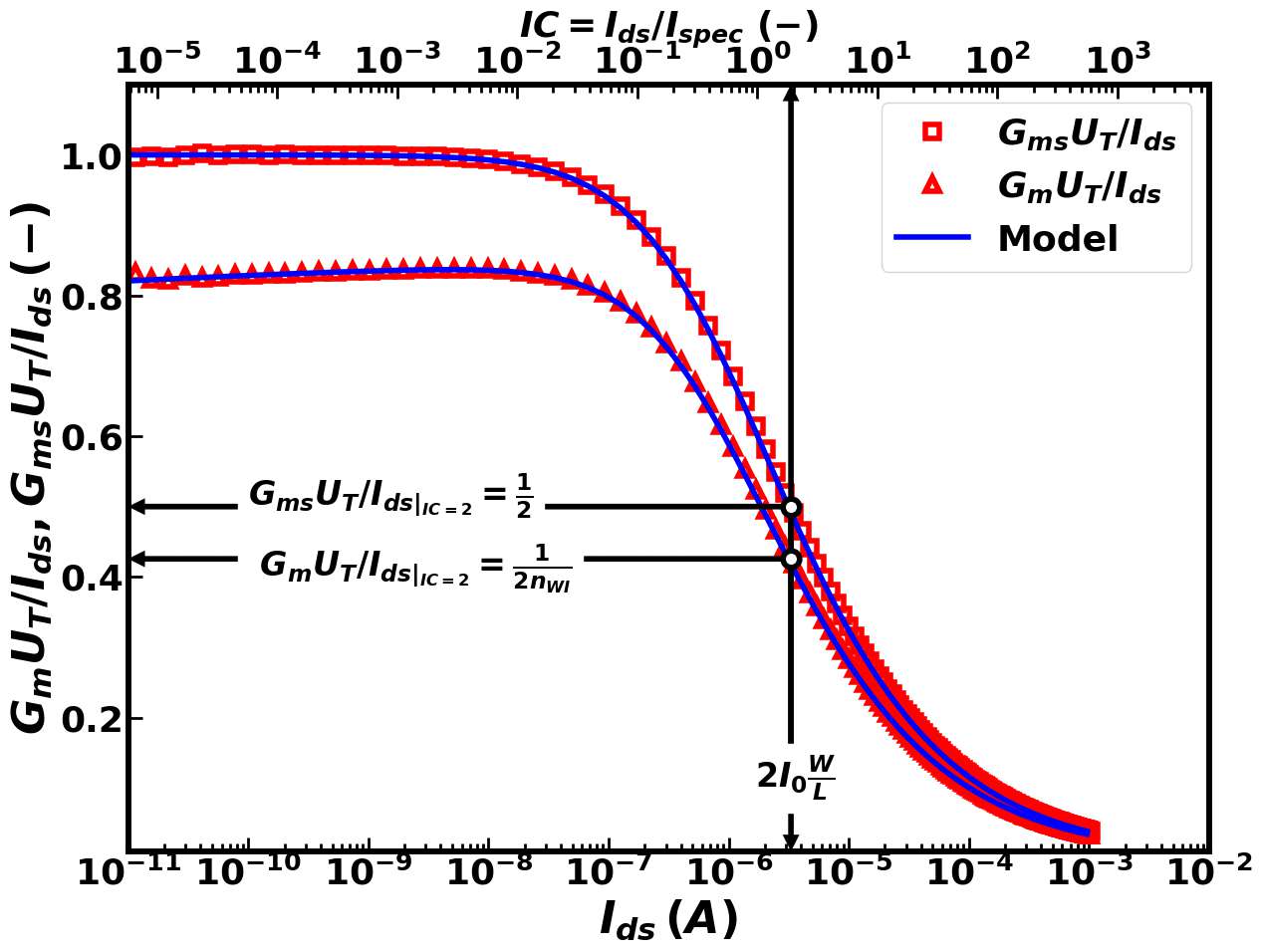} 
	\end{tabular}} \caption{Normalized transconductance-to-current ratio $g_{ms}/{i_d}$ and $g_{m}/{i_d}$ vs. drain current $I_{ds}$ and $IC$ (TCAD: markers, model: lines) to determine
	the specific current $I_{spec}\,=\,I_0\,({W}/{L})$. Here, $IC=2$ 
	corresponds	to the drain current $I_D$ where 50\% of the maximum of $g_{ms}/{i_d}$ or $g_{m}/{i_d}$ in weak inversion is attained. }
\end{figure}

\begin{equation}
\frac{g_{ms}}{i_d}=\frac{n g_{m}}{i_d}|_{sat.}=\frac{1}{1+q_s}=\frac{1}{\frac{1}{2}+\sqrt{\frac{1}{4}+IC}}
\label{gms-ic}
\end{equation}
where $g_{m(s)} = +(-)\partial {I_D} / \partial {V_{G(S)}} \cdot U_T / I_{spec}$ are the normalized gate(source) transconductances. Biasing a transistor at $IC=0.608$ corresponds to pinchoff condition at $g_{m(s)}/i_d=0.701$  (cf. Table I). A certain level of $g_{m(s)}/i_d$, namely 61.8\% of its maximum value in weak inversion, corresponds to $IC=1$ \cite{b5}. As is shown in Fig.~3, $IC=2$ corresponds to 50\% of the maximum of $g_{m(s)}/i_d$ via \eqref{gms-ic} (neglecting the slight bias dependence of the slope factor $n$).

\begin{figure*}[tbph]
	\centerline{%
		\begin{tabular}{cccc}
			\includegraphics[scale=0.415]{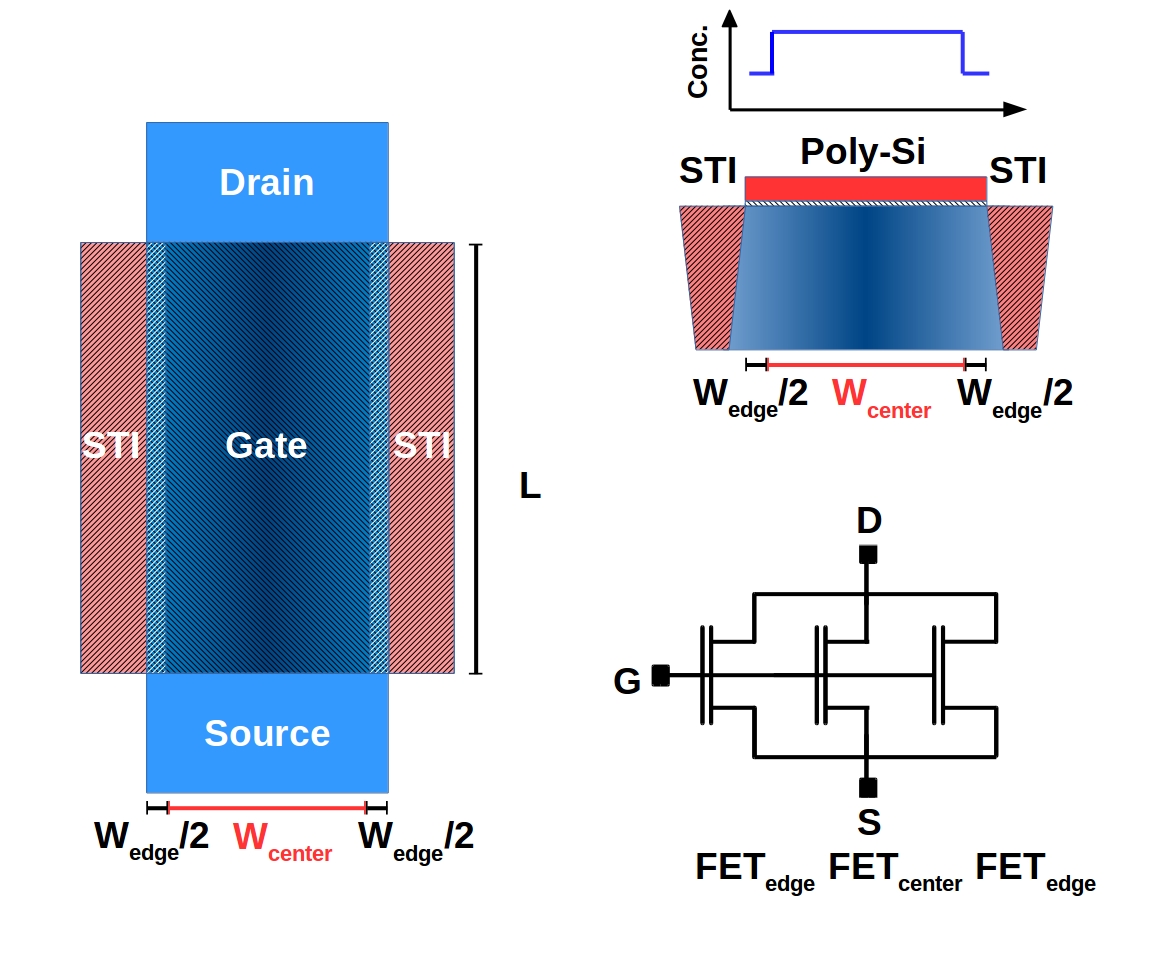}  & 			\includegraphics[scale=0.155]{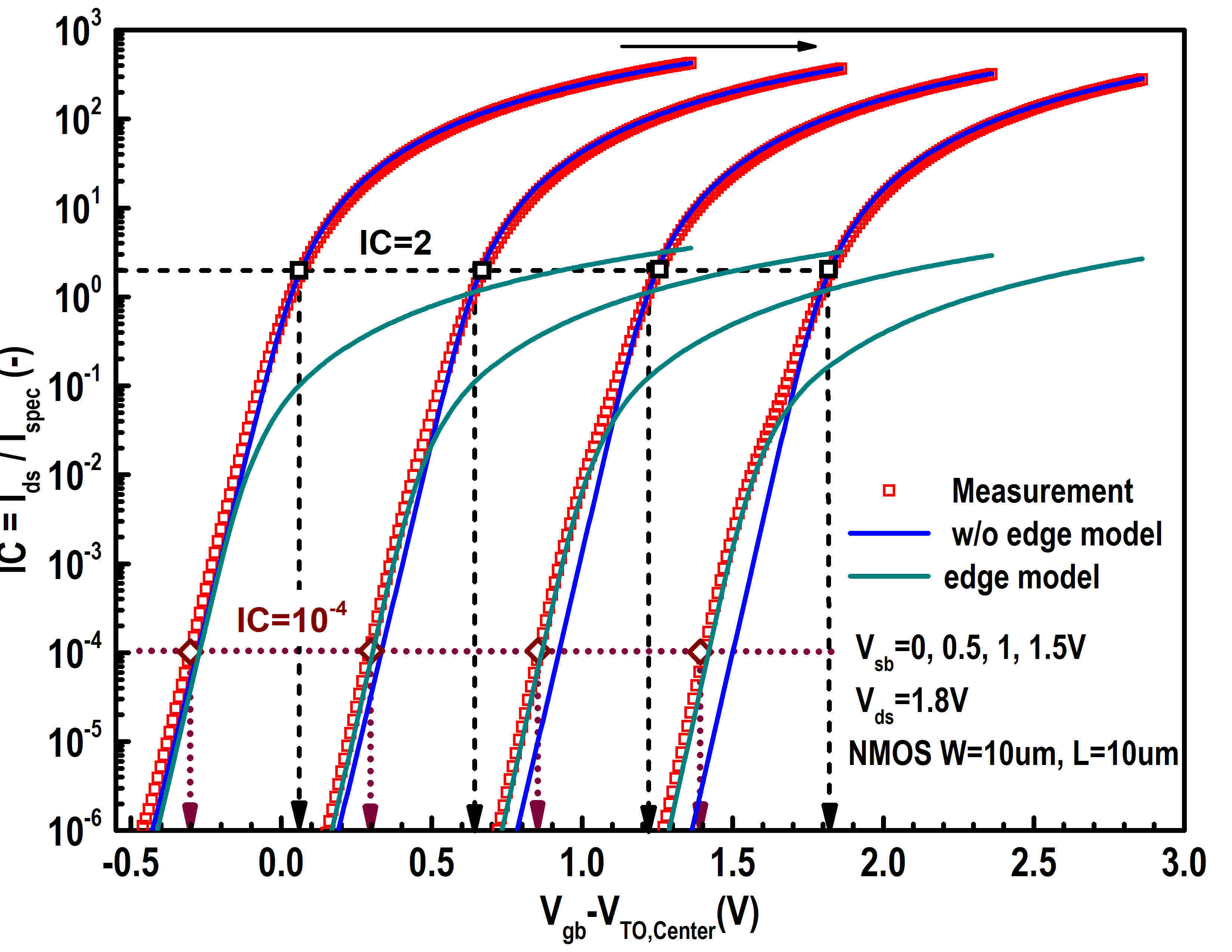}  & \includegraphics[scale=0.155]{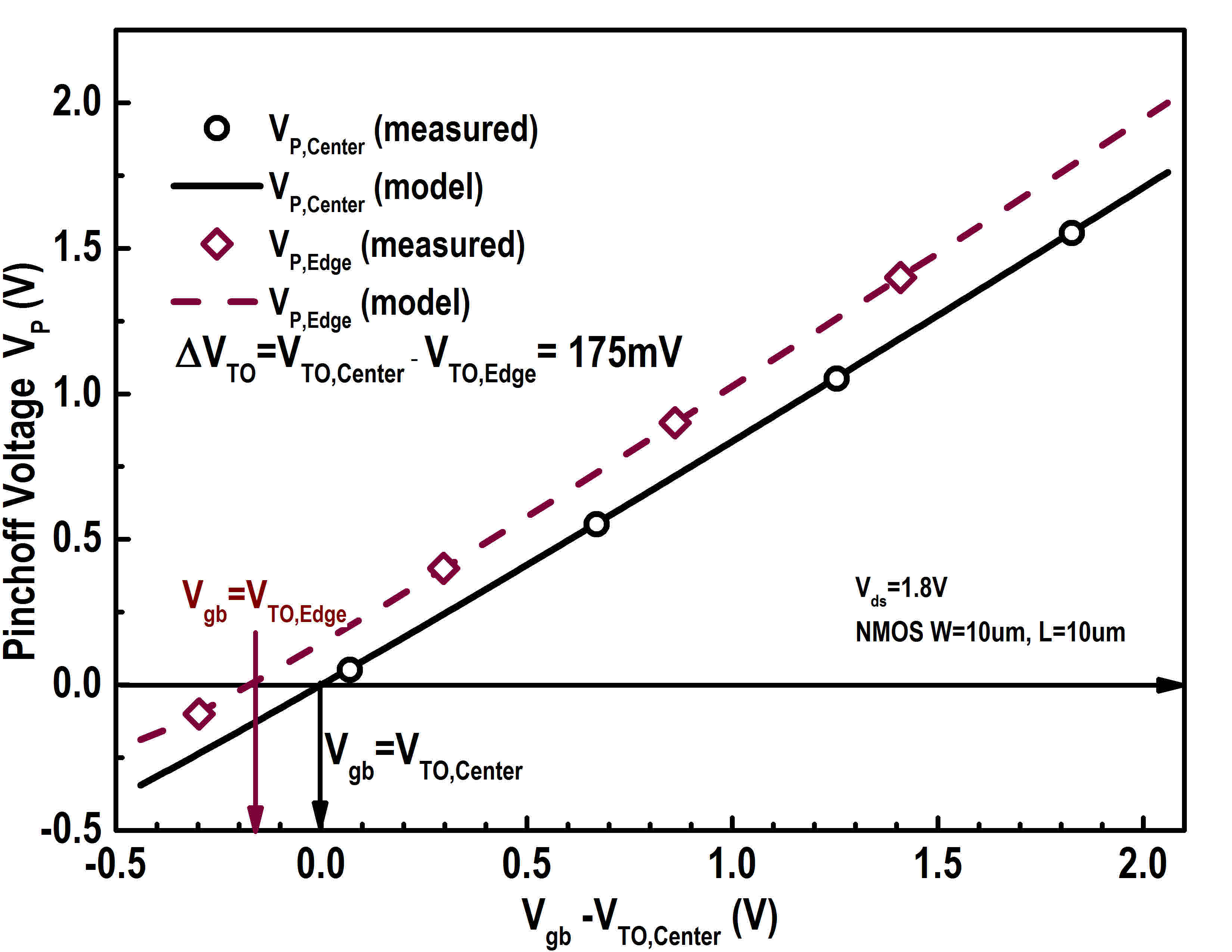} &
            \includegraphics[scale=0.155]{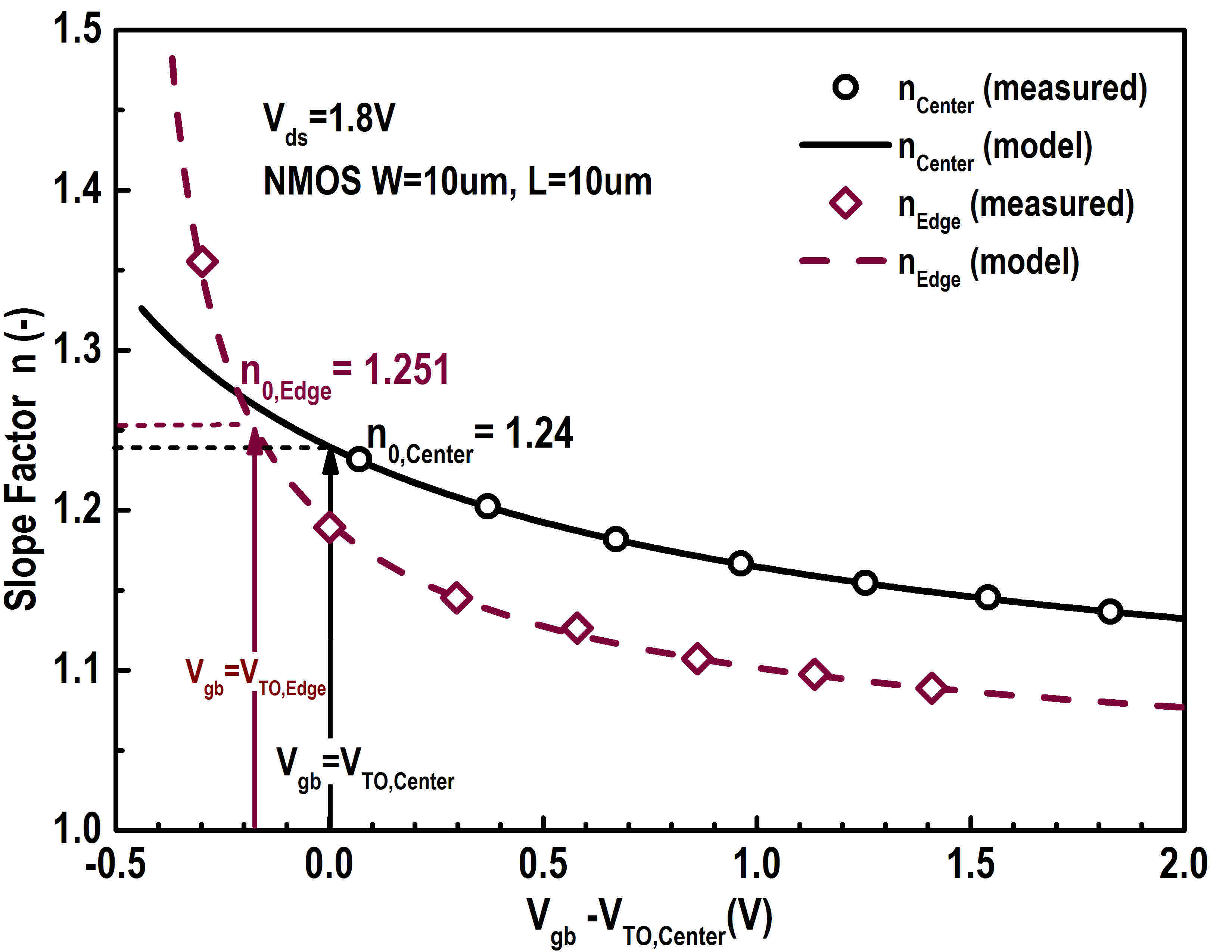}
	\end{tabular}} \caption{Application of the GCC method in presence of edge conduction phenomenon in STI MOSFETs. Width of center and edge transistors sum up to the total width $W=W_{c}+W_{e}$. Transfer characteristics of a wide/long n-MOSFET in saturation (measured data: markers) show edge conduction effect at inversion levels below $IC \approx 1$. Models (full and dashed lines) for the current of the center and edge transistors are shown. A constant current is applied to	characterize the center transistor at $IC=2$, while
	imposing a current criterion $IC=10^{-4}$ corresponds to $IC_e \approx 0.02$ 
	for the edge transistor. $V_P$ and $n$ vs. $V_G$ characteristics illustrate the determination of parameters for center and edge transistors.}
\end{figure*}

\section{Application of GCC to Edge Conduction}
Shallow trench isolation (STI) is the preferred isolation scheme in advanced CMOS technology. As illustrated in Fig.~4, the STI regions at the side edges of the MOSFETs present a lower channel doping concentration due to dopant segregation, and possible thinning of the thin oxide \cite{b12}. As a result, the so-called edge transistors, which operate in parallel to the center transistor, may dominate the subthreshold conduction, strongly impacting on analog circuit operation \cite{b13}. Edge and center transistors have a width of $W_e$ and $W-W_e$, respectively, and the same channel length $L$, hence,
\begin{equation}
\begin{split}
I_{D}&=I_0 \cdot \left[ \frac{W-W_e}{L} \left(i_{f} - i_{r}\right)  + \frac{W_e}{L}  \left(i_{fe} - i_{re}\right) \right]
\label{idtot}
\end{split}
\end{equation}
where the same technology current $I_0$ has been assumed for both components. The current of the edge transistor is obtained using a separate set of equations of the charge-based model \eqref{i-q}-\eqref{vpvg} and \eqref{n} with the edge transistor's parameters, $V_{TOe}$, $\gamma_{e}$, $\Psi_{0e}$. This modeling approach has been used in EKV3 MOSFET model \cite{b14}, and was adopted in BSIM-Bulk \cite{b15}.

The specific current and inversion coefficient of the edge transistor are related to that of the central transistor as,
\begin{equation}
\frac{I_{spec,e}}{I_{spec}} = \frac{W_e} {W-W_e}\,, \quad \quad \frac{IC_e}{IC} = \frac{W-W_e} {W_e}\,.
\label{ispece}
\end{equation}

When applying the GCC method to a transistor with edge conduction effect, a suitable current criterion must be chosen above the level where the hump is apparent, to determine parameters of the central transistor. This could be significantly higher than $IC=2$ used here, particularly in narrow transistors. Conversely, a level well below $IC=0.1$ may be needed to characterize
the edge transistor. For the latter, it is naturally important that the inversion coefficient $IC_e$ \eqref{ispece} is used instead of $IC$ in \eqref{vp-ic}. Fig. 4 illustrates the significant shift in threshold voltage and slope factor $n$ for edge w.r.t. central transistor. Here, the current criterion of $IC=10^{-4}$ corresponds to $IC_e \approx 0.02$ of the edge transistor.

\section{Conclusion}
\label{sec:conclusion}
A new generalized constant current (GCC) method allows for the extraction of MOSFET threshold voltage and other substrate effect-related parameters, from measured drain current at any inversion level from weak to strong inversion. The GCC method is closely related to transconductance-to-current ratio. The method is intricately related to the charge-based model of the MOSFET but may be applied to many other models. A constant current criterion $I_{D,gcc} = I_0 \cdot (W/L) \cdot IC$, where $IC$ defines the desired level of inversion, is typically applied to saturated $I_D-V_G$ characteristics of MOSFETs. Weak and moderate levels of inversion are preferable due to reduced high-field effects. Pinchoff voltage $V_P$ vs. $V_G$, slope factor $n$ vs. $V_G$, and threshold voltage $V_{TH}$ vs. $V_S$ characteristics are obtained, from which $V_{TO}$ and other substrate effect parameters may be determined, irrespectively of the chosen level of inversion $IC$. 
Extracted parameters show a low sensitivity to $IC$, while $V_{TH}$ vs. $V_S$ characteristics are in close agreement with transconductance-to-current ratio change method. As an example, the GCC method is applied to the edge conduction effect in STI-isolated MOSFETs.

\end{document}